\documentclass{webofc}

\usepackage[varg]{txfonts}   
\usepackage{hyperref}
\usepackage{url}
\usepackage{lineno}
\hypersetup{colorlinks=true,citecolor=blue,urlcolor=blue,linkcolor=blue}
\begin{document}
%
\title{New Hypernuclei Measurements from STAR}
%
%

\author{\firstname{Yingjie} \lastname{Zhou}\inst{1,2}\fnsep\thanks{\email{z.yingjie@gsi.de}}(For the STAR Collaboration)
}

\institute{GSI Helmholtzzentrum f\"ur Schwerionenforschung GmbH, Planckstr. 1, 64291 Darmstadt, Germany 
\and 
Institute of Particle Physics and Key Laboratory of Quark \& Lepton Physics (MOE),Central China Normal University, 430079, Wuhan, China
          }

\abstract{Hypernuclei are bound states of hyperons (Y) and nucleons (N). Measurements on their yields can help us investigate their production mechanisms. In particular, the ${}^5_{\Lambda}$He and  $^{4}_{\Lambda}$H(e) are substantially tighter bound compared to the $^{3}_{\Lambda}$H. The large radius of the $^{3}_{\Lambda}$H leads to suppression in coalescence models, but not in the thermal model where the size of the nucleus does not play a role. As such, studying the $A=3$--5 hypernuclei yields allow us to extract information on the effects of hypernuclear binding on hypernuclei production in heavy-ion collisions. 

In these proceedings, we present measurements of ${}^5_{\Lambda}$He yields in Au+Au collisions at $\sqrt{s_{\rm NN}}=3.0$ GeV, $^{4}_{\Lambda}$H(e) yields in Au+Au collisions at $\sqrt{s_{\rm NN}}=3.0$--$4.5$ GeV, and $^{3}_{\Lambda}$H yields in Au+Au collisions at $\sqrt{s_{\rm NN}}=3$--$27$ GeV. Results on the directed flow of hypernuclei are also reported. The physics implications of these measurements are discussed. }
 
\maketitle
\section{Introduction}
\label{intro}
Hypernuclei, as bound systems of nucleons and hyperons, offer a unique laboratory to investigate hyperon--nucleon (Y--N) and even hyperon--nucleon--nucleon (Y--N--N) interactions in dense nuclear matter. These studies are particularly relevant to addressing the so-called hyperon puzzle in neutron stars~\cite{Lonardoni:2014bwa}. A central question is whether measurements of hypernuclei production can provide meaningful constraints on these in-medium interactions. Therefore, a detailed understanding of the production mechanisms of hypernuclei is essential for utilizing them as effective probes of the Y--N interaction.

Several theoretical models have been developed to describe hypernuclei production. The thermal (statistical hadronization) model assumes that all particles are produced in chemical equilibrium, whereas coalescence models propose that (hyper)nuclei are formed by the clustering of nucleons after kinetic freeze-out. Recent measurements of light nuclei yields have challenged the thermal model: while the $d/p$ is fairly well described, the $t/p$ and $^3\mathrm{He}/p$ ratios are significantly overestimated~\cite{STAR:2023uxk, ALICE:2022veq}.

\section{STAR Beam Energy Scan II and hypernuclei reconstruction}
The production of hypernuclei in heavy-ion collisions is expected to increase at lower beam energies due to the higher baryon density~\cite{Andronic:2010qu,Steinheimer:2012tb}. The STAR Beam Energy Scan II program, spanning collision energies from $\sqrt{s_{\rm NN}} = 3.0$ to $27.0$~GeV, 3.0-7.7 GeV in fixed-target (FXT) mode and 7.7-27 GeV in collider mode, offers a unique opportunity for systematic studies of hypernuclei production. Measurements of ${}^{3}_{\Lambda}\mathrm{H}$ and ${}^{4}_{\Lambda}\mathrm{H}$ yields have been performed in Au+Au collisions at $\sqrt{s_{\rm NN}} = 3.0$~GeV, utilizing a dataset of 258 million events collected in 2018~\cite{STAR:2021orx}. In 2021, STAR recorded 2 billion events at $\sqrt{s_{\rm NN}} = 3.0$~GeV, enabling the first measurement of ${}^{5}_{\Lambda}\mathrm{He}$ yield and flow in heavy-ion collisions. The observation of ${}^{5}_{\Lambda}\mathrm{He}$ is particularly noteworthy, as it represents the heaviest hypernucleus measured to date in such collisions.

Hypernuclei are reconstructed via their weak decays, such as ${}^{5}_{\Lambda}\mathrm{He} \rightarrow p + {}^{4}\mathrm{He} + \pi^{-}$. Daughter particles are identified using their ionization energy loss ($dE/dx$) measured in the Time Projection Chamber. Efficiency corrections are obtained from data-driven GEANT simulations. To accurately model the three-body decay phase space, the Dalitz distribution extracted from data is used to weight the simulated decays, ensuring precise reproduction of the observed kinematic distributions.
\section{Results and Discussion}
\subsection{Particle Yields}
\label{sec-1}
\begin{figure}[h]
\centering
\sidecaption
\includegraphics[width=4cm,clip]{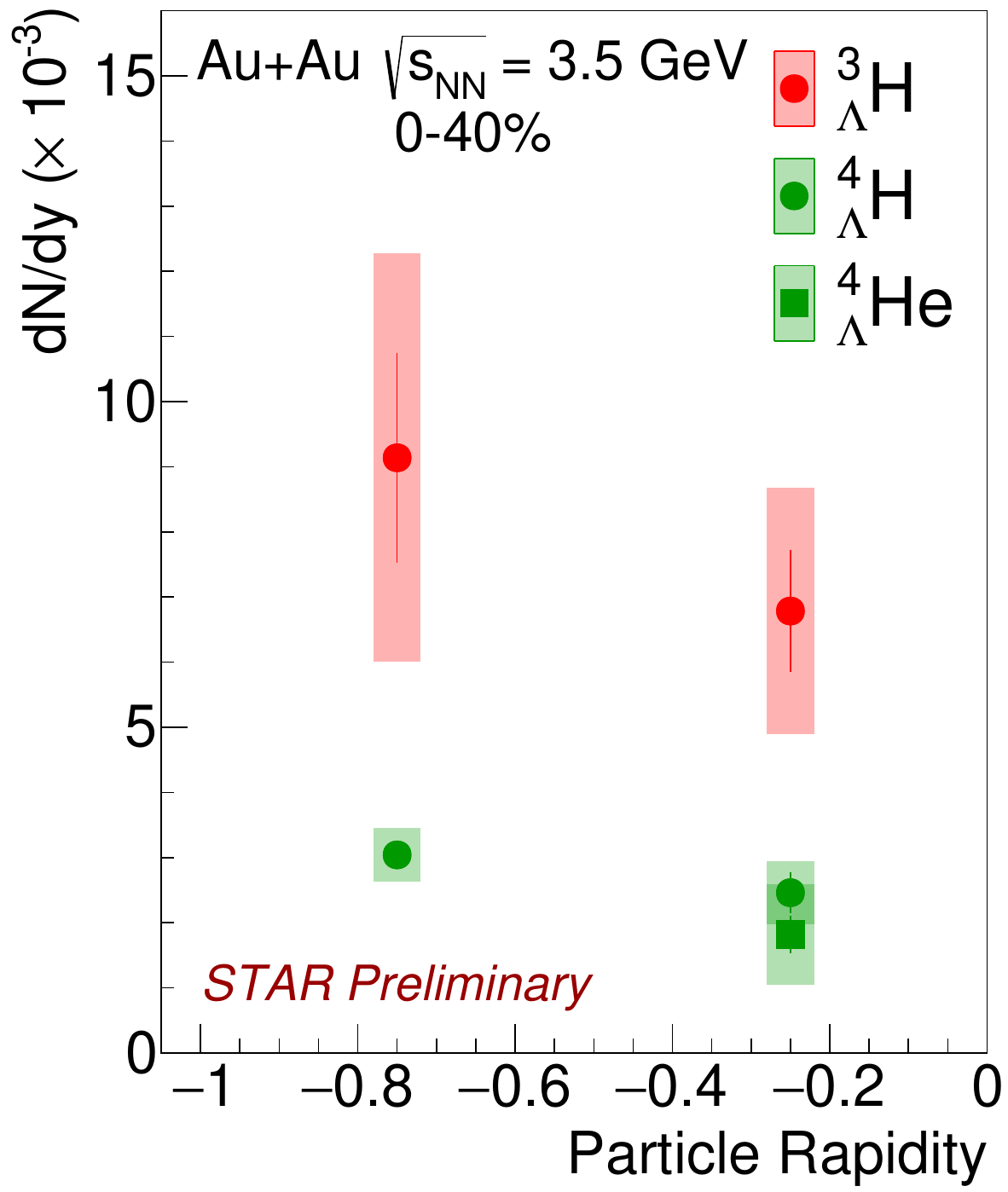}
\includegraphics[width=4cm,clip]{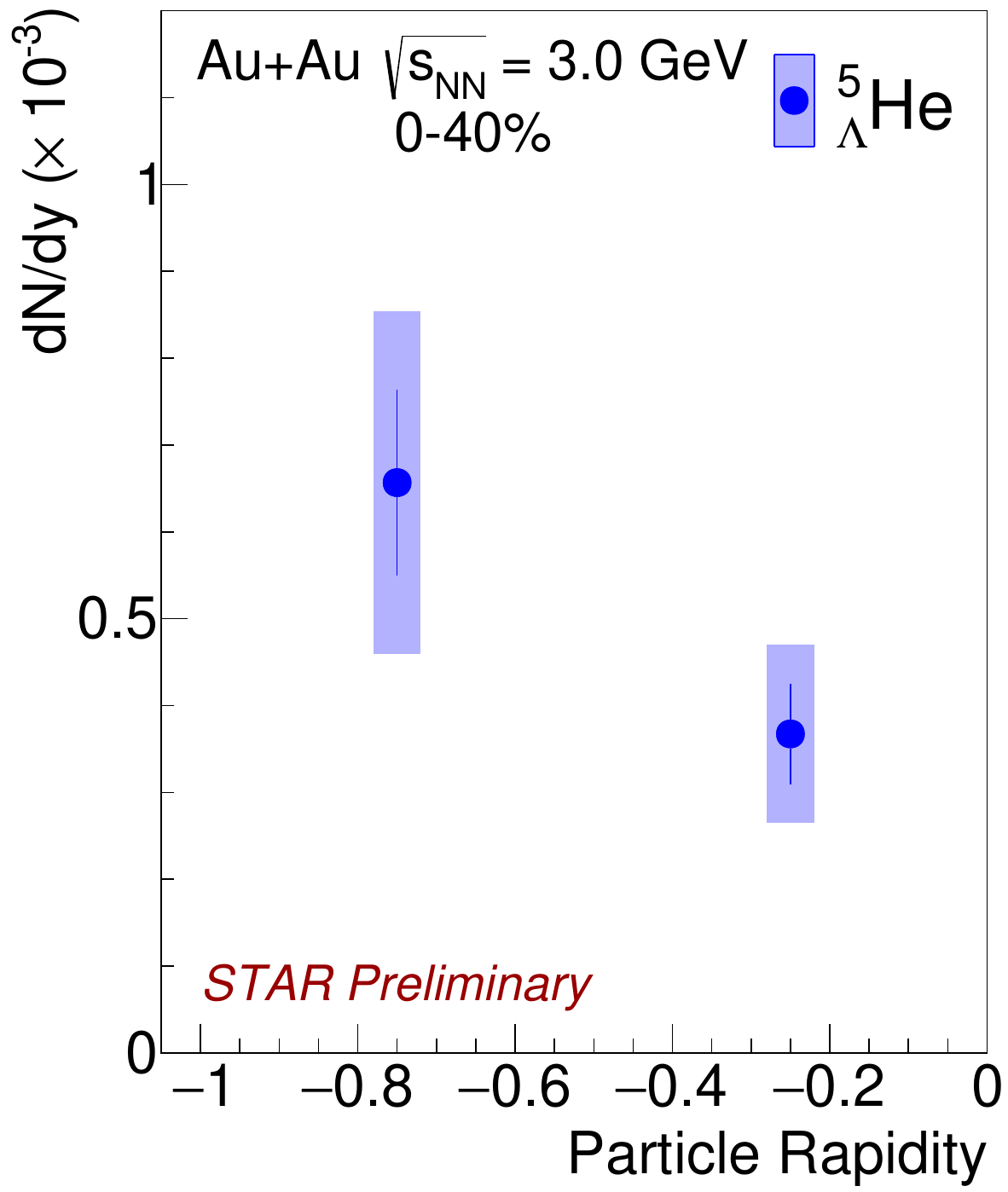}
\caption{Rapidity yield distributions of hypernuclei ${}^{3}_{\Lambda}\mathrm{H}$, ${}^{4}_{\Lambda}\mathrm{H}$, ${}^{4}_{\Lambda}\mathrm{He}$ at $\sqrt{s_{\rm NN}}=3.5$~GeV, and ${}^{5}_{\Lambda}\mathrm{He}$ at $\sqrt{s_{\rm NN}}=3.0$~GeV in 0--40\% Au+Au collisions.}
\label{fig:3gevdndy}       
\end{figure}

Figure~\ref{fig:3gevdndy} shows the rapidity distributions for hypernuclei ${}^{3}_{\Lambda}\mathrm{H}$, ${}^{4}_{\Lambda}\mathrm{H}$, ${}^{4}_{\Lambda}\mathrm{He}$ at $\sqrt{s_{\rm NN}}=3.5$~GeV, and ${}^{5}_{\Lambda}\mathrm{He}$ at $\sqrt{s_{\rm NN}}=3.0$~GeV in 0--40\% Au+Au collisions. Significant yields are observed at target rapidity (y=$-1.05$) for all species, including ${}^{5}_{\Lambda}\mathrm{He}$, suggesting that spectator matter plays an increasingly important role in their production at low energies~\cite{Botvina:2011jt}.

\begin{figure}[h]
\centering
\includegraphics[width=5.cm,clip]{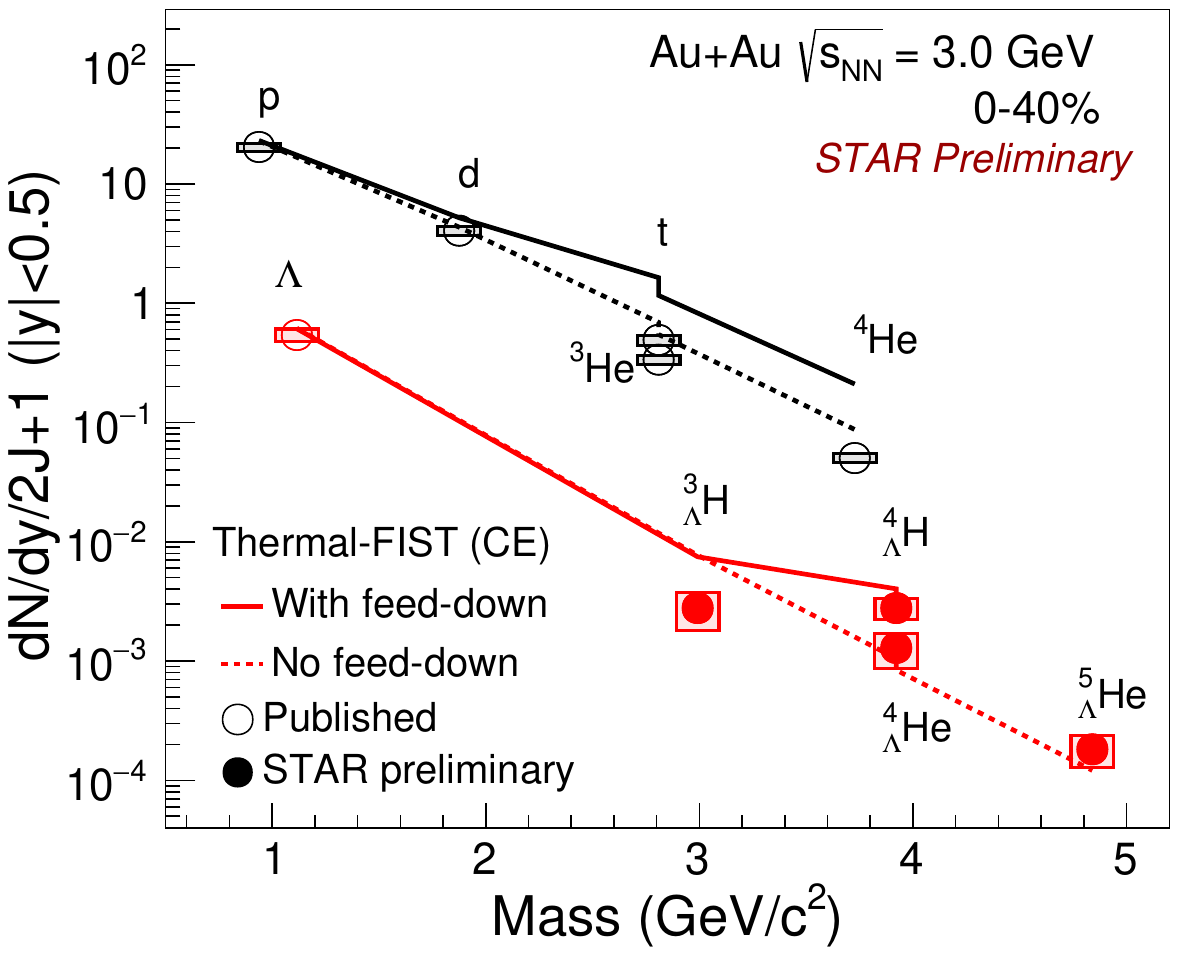}
\includegraphics[width=5 cm,clip]{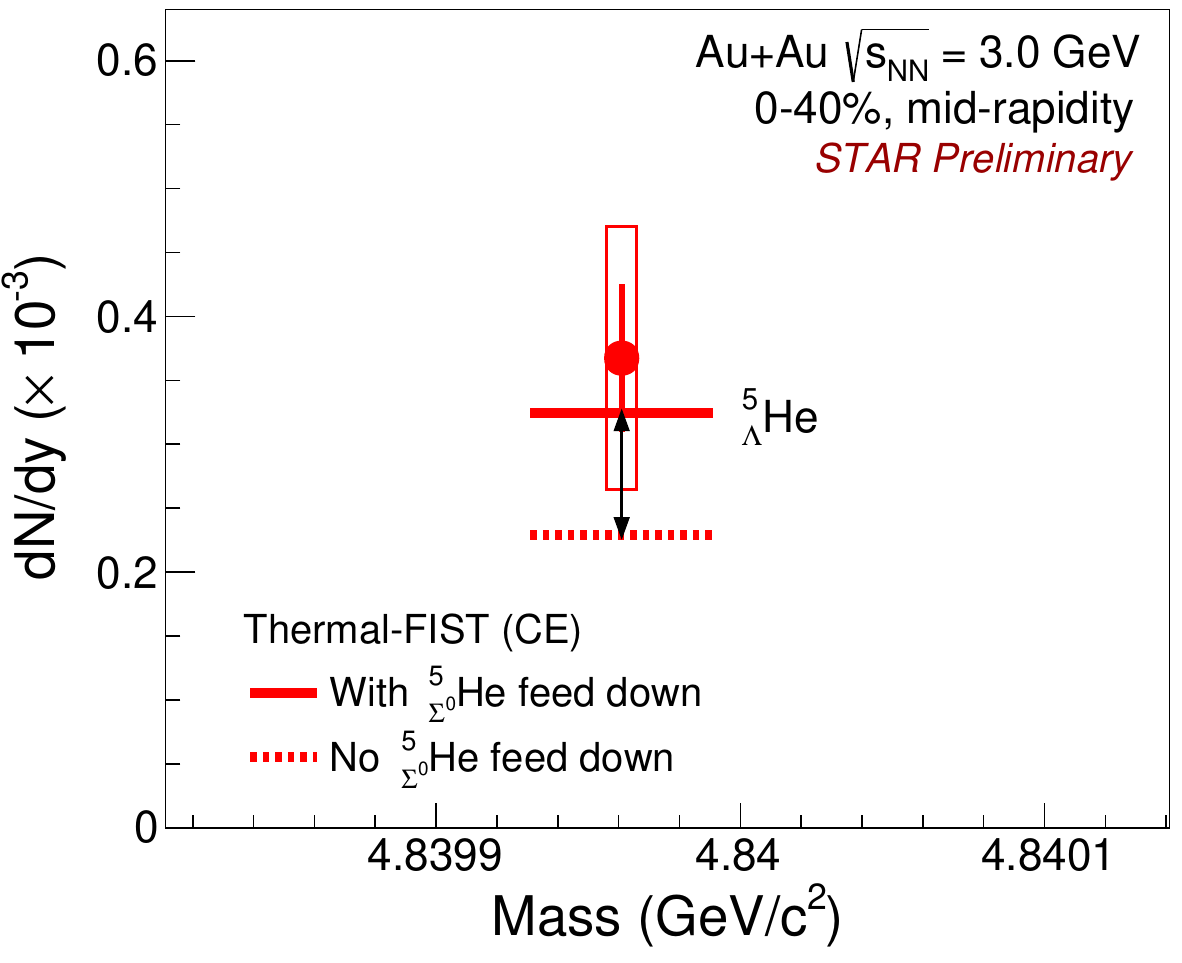}
\caption{Left: Measured mid-rapdity yields of light (hyper)nuclei in Au+Au collisions at $\sqrt{s_{\rm NN}}=3.0$~GeV, scaled by their spin degeneracy factor $(2J+1)$, as a function of mass number $A$, compared to thermal model calculations. Right: Comparison of measured and thermal model predicted yields for $^5_\Lambda$He, including contributions from unstable nuclei. Data from~\cite{STAR:2023uxk, STAR:2024znc, STAR:2021orx}.}
\label{fig:dndySpin}  
\end{figure}  

\begin{figure}[h]
\centering
\sidecaption
\includegraphics[width=5.cm,clip]{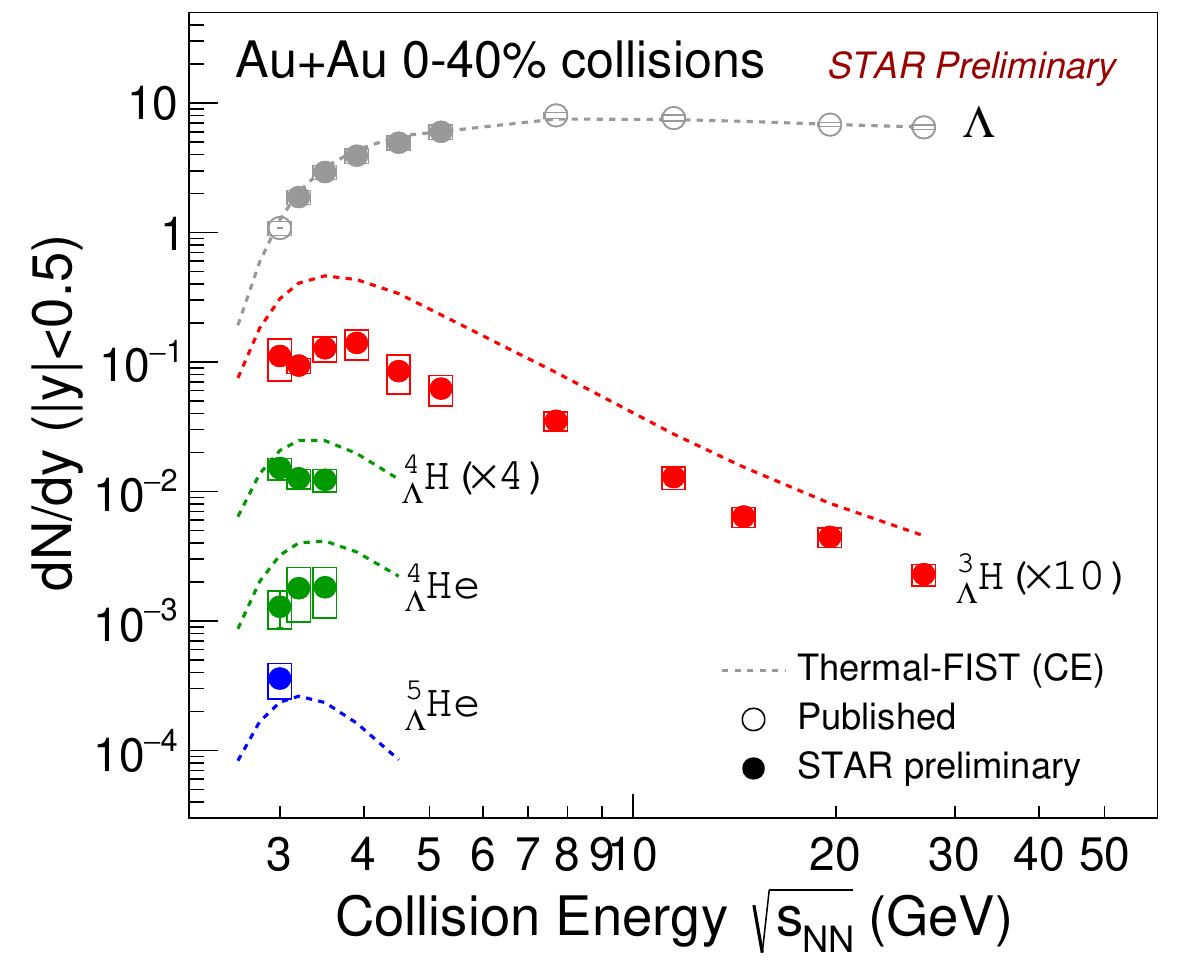}
\caption{Energy dependence of the measured mid-rapidity yields (dN/dy) of $\Lambda$, ${}^{3}_{\Lambda}\mathrm{H}$, ${}^{4}_{\Lambda}\mathrm{H}$, ${}^{4}_{\Lambda}\mathrm{He}$, and ${}^{5}_{\Lambda}\mathrm{He}$ in 0--40\% Au+Au collisions. The dashed line represents thermal model predictions~\cite{Vovchenko:2015idt}.}
\label{fig:dndySnn}      
\end{figure}

The left panel of Fig.~\ref{fig:dndySpin} shows the measured yields of light (hyper)nuclei at $\sqrt{s_{\rm NN}}=3.0$~GeV at mid-rapidity, scaled by their spin degeneracy factor $(2J+1)$, as a function of mass number $A$. These results are compared to thermal model calculations~\cite{Vovchenko:2015idt}, which predict an exponential decrease of yield/$(2J+1)$ with increasing $A$. The dashed line represents the thermal model without feed-down from unstable nuclei, while the solid line includes such contributions. The thermal model with feed-down overestimates light-nuclei yields and overpredicts $^4_{\Lambda}\mathrm{H}$ and $^4_{\Lambda}\mathrm{He}$ when excited-state contributions are included(e.g., $^4_{\Lambda}\mathrm{H}^*$ and $^4_{\Lambda}\mathrm{He}^*$~\cite{A1:2016nfu}). In contrast, thermal model predictions lie slightly below the measured $^5_{\Lambda}\mathrm{He}$ yield, which hints at possible contributions from $^5_{\Sigma^{0}}\mathrm{He} \rightarrow {}^5_{\Lambda}\mathrm{He} + \gamma$~\cite{Johnstone:1981ih}. As shown in the right panel of Fig.~\ref{fig:dndySpin}, including this contribution leads to improved agreement between the thermal model and experimental data. Figure~\ref{fig:dndySnn} presents the energy dependence of dN/dy for $\Lambda$ and hypernuclei at mid-rapidity (|y|$<0.5$) in 0--40\% Au+Au collisions across a range of $\sqrt{s_{\rm NN}}$ from 3 to 27~GeV. The thermal model describes the overall trend, but overestimates the yields of $^3_\Lambda$H, $^4_\Lambda$H, and $^4_\Lambda$He, while slightly underestimating $^5_\Lambda$He.

\subsection{Collectivity}
\label{sec-2}
\begin{figure}[h]
\centering
\includegraphics[width=5cm,clip]{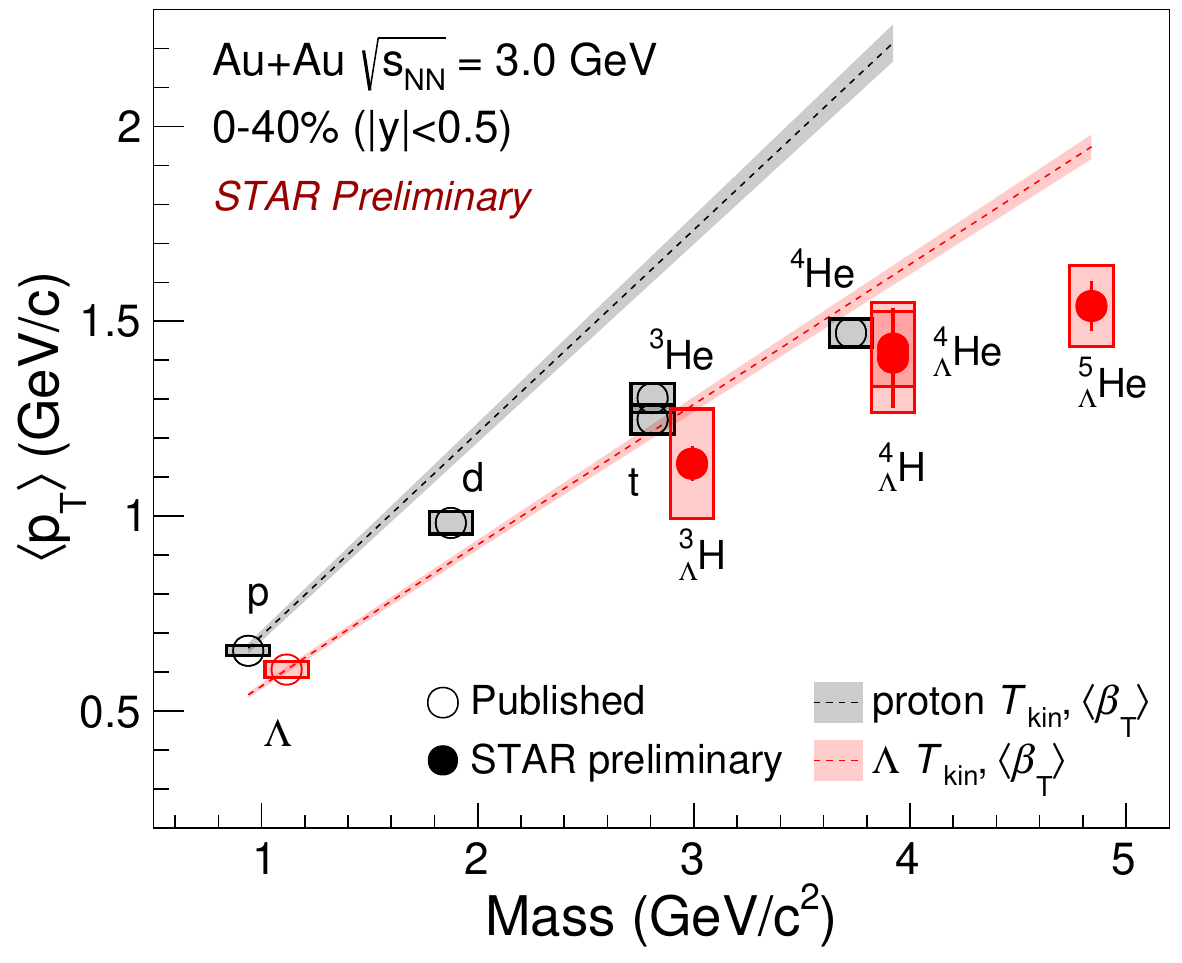}
\includegraphics[width=5cm,clip]{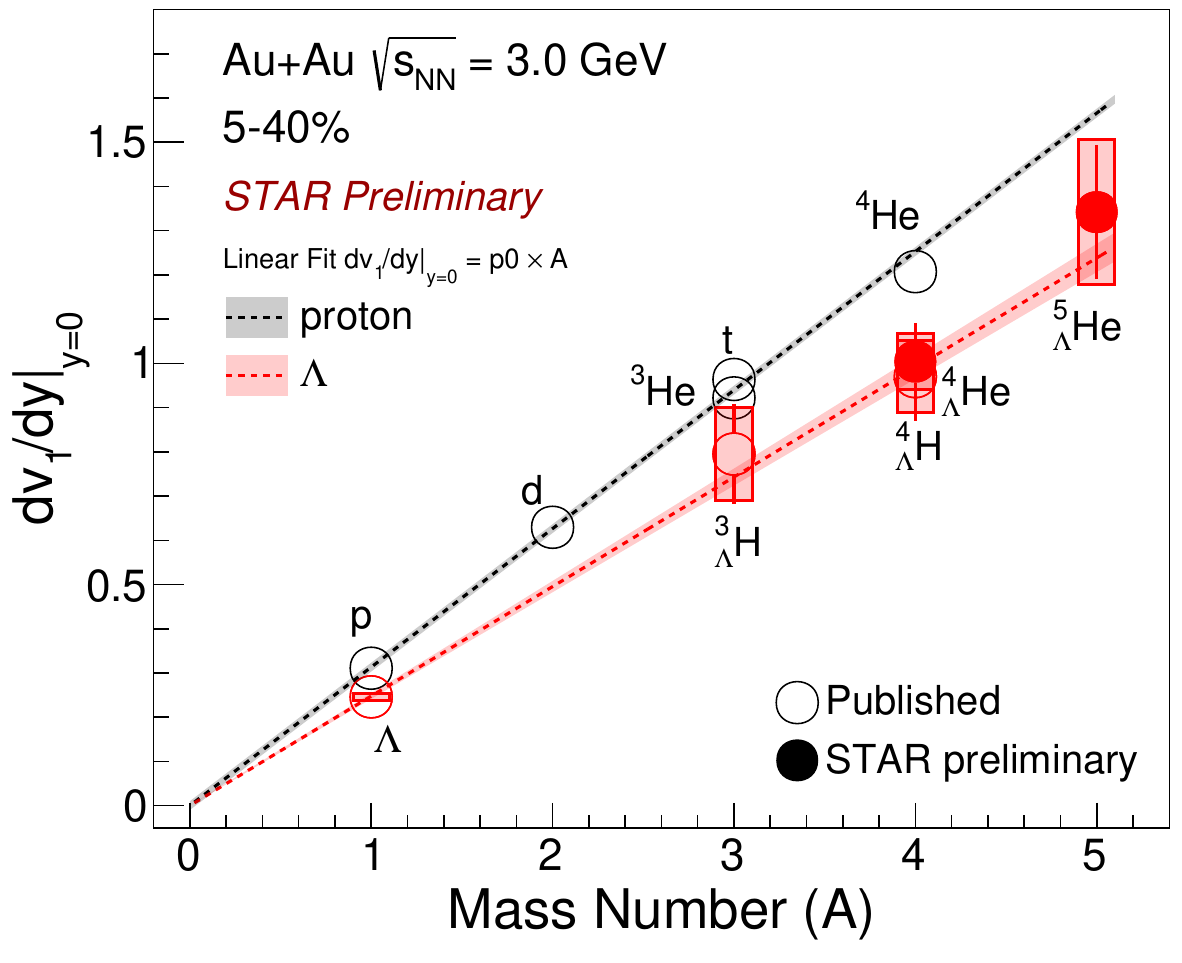}
\caption{Left: Mass dependence of the mid-rapidity $\langle p_\mathrm{T} \rangle$ for light (hyper)nuclei from $\sqrt{s_{\rm NN}} = 3.0$~GeV in 0--40\% Au+Au collisions. The symbols represent measurements, while the lines represent hydrodynamic-inspired Blast-Wave model calculations. The gray line shows the prediction using proton freeze-out parameters, while the red line shows the prediction using $\Lambda$ parameters. Right: Directed flow slope $dv_{1}/dy$ at mid-rapidity for light (hyper)nuclei as a function of mass number in $\sqrt{s_{\rm NN}}=3.0$~GeV 5--40\% Au+Au collisions. Black and red bands indicate linear fits to the light (hyper)nuclei, respectively.}
\label{fig:3gevmeanpt}       
\end{figure}
Figure~\ref{fig:3gevmeanpt} (left) shows the mean transverse momentum, $\langle p_\mathrm{T} \rangle$, of light (hyper)nuclei at mid-rapidity in $\sqrt{s_{\rm NN}} = 3.0$~GeV Au+Au collisions. The data are compared to hydrodynamic-inspired Blast-Wave model predictions~\cite{Liu:2024ygk}, which assume thermal emission from an expanding source with a common kinetic freeze-out temperature ($T_{\text{kin}}$) and average transverse flow velocity ($\langle \beta_\text{T} \rangle$). The gray and red lines represent the Blast-Wave predictions using the freeze-out parameters for protons and $\Lambda$s, respectively. The measured $\langle p_\mathrm{T} \rangle$ values for light (hyper)nuclei systematically fall below these curves, indicating a clear deviation from thermal expectations. The right panel of Fig.~\ref{fig:3gevmeanpt} presents the mid-rapidity $v_1$ slope for light (hyper)nuclei at $\sqrt{s_{\rm NN}}=3.0$~GeV. The black and red bands correspond to linear fits for light (hyper)nuclei, respectively, as a function of mass number. The results demonstrate a clear scaling of $v_1$ slope with mass, indicating that heavier particles exhibit larger directed flow. This mass scaling is consistent with coalescence model expectations, where the collective motion of constituent nucleons is reflected in the flow of the composite nuclei.

\section{Summary}
In summary, we present the first measurement of ${}^{5}_{\Lambda}\mathrm{He}$ hypernuclei yields and flow in Au+Au collisions at $\sqrt{s_{\rm NN}}=3.0$~GeV, alongside results for other light (hyper)nuclei. Our results shows that the thermal model overestimates the yields of $A\leq4$ hypernuclei, while slightly underestimating the ${}^{5}_{\Lambda}\mathrm{He}$ yield, providing a first hint for feed-down contributions from $\Sigma$ hypernuclei. Both light (hyper)nuclei exhibit freeze-out conditions distinct from those of bulk particles such as protons and $\Lambda$s. An approximate atomic mass number scaling is observed in the measured mid-rapidity $v_1$ slopes of light (hyper)nuclei. All measurements consistently point to the coalescence production mechanism for light (hyper)nuclei.

\section{Acknowledgement}
This work was supported by the National Natural Science Foundation of China under Grant No. 12375134, the National Key Research and Development Program of China (Grant No. 2024YFE0110103 and 2024YFA1611003), and the Fundamental Research Funds for the Central Universities (Grant No. CCNU25JCPT017), and the FAIR Fellowship and Associate Program of GSI Helmholtzzentrum für Schwerionenforschung, Darmstadt, Germany.

\bibliography{ref} 

\begin{thebibliography}{12}

\bibitem{Lonardoni:2014bwa}
D. Lonardoni {\textit{et~al.}}, Phys. Rev. Lett. \textbf{114}, 092301 (2015)

\bibitem{STAR:2023uxk}
M. Abdulhamid {\textit{et~al.}} {[STAR Collaboration]}, Phys. Rev. C
  \textbf{110}, 054911 (2024)

\bibitem{ALICE:2022veq}
S. Acharya {\textit{et~al.}} {[ALICE Collaboration]}, Phys. Rev. C
  \textbf{107}, 064904 (2023)

\bibitem{Andronic:2010qu}
A. Andronic {\textit{et~al.}}, Phys. Lett. B \textbf{697}, 203 (2011)

\bibitem{Steinheimer:2012tb}
J. Steinheimer {\textit{et~al.}}, Phys. Lett. B \textbf{714}, 85 (2012)

\bibitem{STAR:2021orx}
M. Abdallah {\textit{et~al.}} {[STAR Collaboration]}, Phys. Rev. Lett.
  \textbf{128}, 202301 (2022)

\bibitem{Botvina:2011jt}
A.S. Botvina {\textit{et~al.}}, Phys. Rev. C \textbf{84}, 064904 (2011)

\bibitem{STAR:2024znc}
M.I. Abdulhamid {\textit{et~al.}} {[STAR Collaboration]}, JHEP \textbf{10}, 139
  (2024)

\bibitem{Vovchenko:2015idt}
V. Vovchenko {\textit{et~al.}}, Phys. Rev. C \textbf{93}, 064906 (2016)

\bibitem{A1:2016nfu}
F. Schulz {\textit{et~al.}} {[A1 Collaboration]}, Nucl. Phys. A \textbf{954},
  149 (2016)

\bibitem{Johnstone:1981ih}
J. Johnstone {\textit{et~al.}}, J. Phys. G \textbf{8}, L105 (1982)

\bibitem{Liu:2024ygk}
D.N. Liu {\textit{et~al.}}, Phys. Lett. B \textbf{855}, 138855 (2024)

\end{thebibliography}
\end{document}